\documentstyle[12pt]{article}
\newcommand{\inbrace}{\rangle^{\rm in}}
\newcommand{\outbrace}{{}^{\rm out}\langle}
\newcommand{\invac}{| 0 \rangle}

\newcommand{\Pbar}{ {\overline P} }
\newcommand{\FF}{F^{BC}_{J^a_L}}
\newcommand{\topi}{\frac{\theta}{2\pi i}}
\begin{document}

\setcounter{page}{1}

\rightline{Preprint UFIFT-HEP-94-15}
\rightline{hep-th/9410243}

\vskip.5in

\centerline{\large {\bf Light Front Hamiltonian for Transverse Lattice
QCD\footnote{Based on a lecture given
at the ``Theory of Hadrons and Light-front QCD'' workshop
in Zakopane, Poland, August 1994.}
}}
\vskip.3in
\centerline{ Paul A.
Griffin\footnote{e-mail: pgriffin@phys.ufl.edu}
}
\centerline{ Institute for Fundamental Theory, University of Florida}
\centerline{ Gainesville, FL 32611, USA}

\vskip.3 in
\centerline{\bf Abstract}
\vskip.1in
A calculational framework for determining masses of low
lying hadrons using light front quantization is
discussed.  The method is based upon four theoretical tools:
discrete light cone quantization, which has been very successful
in $1+1$ dimensional models, a projector Monte Carlo method to extract
low lying state data with 2 additional transverse dimensions,
the transverse spatial lattice Hamiltonian of QCD,
and exact form factors of the $SU(3)_L\otimes SU(3)_R$ symmetric
$1+1$ dimensional non-linear sigma model (NLSM).  How these tools are
to be put together to provide a description of hadrons
is the main topic of this lecture.  I also focus
on the NLSM form factors, which are given new
physical relevance via this picture of QCD.
\vskip.1in
\newpage

\vskip.1in
{\bf 1. Introduction}
\vskip.1in

I would like to suggest an alternative framework for the
calculation of hadron masses.
This framework is built upon
the method of discrete light cone (front) quantization DLCQ[1],
which combines the natural truncation of physical states on the
light-front with a numerical cutoff in the number of states.  The light
front quantization guarantees that physical states have positive
light front momentum $p^+$, and DLCQ splits the total $p^+$ into discrete
bits.  The number of states in the Hilbert space is given by the number
of possible ways of partitioning the bits of momenta of the particles
in the theory.  The light front Hamiltonian $p^-$ commutes with $p^+$
and is inversely proportional to the bit size, $p^- \sim M^2/p^+$,
so a mass spectrum calculation of the lowest states becomes
tractable.  With this method, the particle data booklet of
$1+1$ dimensional QCD with dynamical
fermions can filled out using only a modern personal computer[2].

Extending this analysis to $3+1$ dimensions in a straightforward
fashion has been unsuccessful so far,  primarily
because of strong mixing of high and low momentum particles.
I propose to use a spatial lattice method to regulate this problem,
much like in ordinary lattice gauge theory.  In this context
however, the lattice is two dimensional,
and the full light front Hamiltonian is decomposed into
($1+1$)-D~longitudinal, 2-D~transverse,
and mixed parts of the theory. One does not
need to diagonalize the entire Hamiltonian to study properties of the lowest
states; diagonalization via DLCQ of
a small subset of the Hamiltonian which includes only nearest or next to
nearest neighbor transverse
interactions is sufficient.  This reduces both the computer time and
memory requirements of a full diagonalization problem.  The true ground
states of the theory are obtained by projecting out the excited state
components of a trial wave-function.  This has been formulated and
successfully applied to the $\phi^4$ theory in $2+1$ dimensions
by Burkardt[3].

The classical action for QCD on a transverse lattice
was given by Bardeen and Pearson a long time ago[4].
Introduce link fields $U_{x_{\perp}}$,  which
are scalars with respect to the two continuous space-time
coordinates perpendicular to the lattice.
In light cone
gauge\footnote{
The `zero mode' issue associated with this gauge is
neglected in this lecture.}
$A^- =0$, and after eliminating $A^+$ via the Gauss
constraint, the action $A = \int dx^+ dx^-
\sum_{x_\perp}{\cal L}$, where ${\cal L} = {\cal L}_{\rm LL} +
{\cal L}_{\rm LT} + {\cal L}_{\rm TT}$, where ${\rm L}$ denotes longitudinal
and ${\rm T}$ denotes transverse,
\begin{eqnarray}
{\cal L}_{\rm LL} &=&
\frac{1}{g_1^2} \sum_{\alpha = 1,2} {\rm tr} (\partial_+ U_{x_\perp,\alpha}
\partial_-  U_{x_\perp,\alpha}^\dagger ) \ , \\
{\cal L}_{\rm LT} &=&
\frac{1}{g_2^2 a^2}\int dy^-
|x^- - y^-|J_- (x^-)\cdot J_- (y^-) \ , \\
{\cal L}_{\rm TT} &=&
\frac{1}{g_3^2 a^2}\sum_{\alpha\beta}{\rm tr} \big(
U_{x_\perp,\alpha} U_{x_\perp + \hat{\alpha},\beta}
U^\dagger_{x_\perp +\hat{\beta}, \alpha } U^\dagger_{x_\perp,\beta}\big)
 \ .
\end{eqnarray}
Eqn. (1) denotes the pure longitudinal part of the action, and is
a set of decoupled two-dimensional non-linear sigma models (NLSM).  The
non-linear constraints are ${\rm det}\, U = 1$ and $U^\dagger = U^{-1}$
for $SU(N)$ theory.  Eqn.~(2) denotes longitudinal gluon exchange
and contains both local and nearest neighbor type interactions.
The pure glue non-Abelian current is given by
$ J^a_- = i \sum_{\alpha} {\rm tr} [ T^a (U^\dagger_{x_\perp,\alpha}
\partial_- U_{x_\perp,\alpha} + U^\dagger_{x_\perp-\hat{\alpha},\alpha}
\partial_- U_{x_\perp-\hat{\alpha},\alpha})]$.
Eqn.~(3) is the nearest neighbor transverse plaquette interaction
which becomes
$F_{\alpha\beta}^2$ in the continuum limit.  The parameter $a$ is
the transverse lattice spacing.

Bardeen, Pearson and Rabinovici performed a DLCQ calculation of the
glueball spectrum in the large $N$ limit[4] by
approximating the NLSM part of the Lagrangian eqn.~(1) with a complex
linear matrix field and adding potential terms to mimic the non-linear
constraints.  However for finite $N$, this scheme is
problematic because the number of induced potential terms
(suppressed by powers of $1/N$ in the large $N$ calculation)
is infinite.  In addition, DLCQ is not
able to describe behavior of systems about phase transition
critical points very well without addressing the `zero mode'
problems of the vacuum, i.e.~the coherent state of the vacuum
in the new phase
is made out of infinitely many Fock state particles,
and is difficult to simulate numerically.

I propose to solve the NLSM part of the theory
exactly, by calculating matrix elements of the
asymptotic states $P_i(\theta_i)$ with operators in the interacting
quantum Hamiltonian, where $\theta = \ln {(P^+ /m)}$ is the rapidity
of state
$P$ with mass $m$.
The relevant form factors of the NLSM are
\begin{eqnarray}
&&\langle J_- (0) | P_1 (\theta_1) P_2 (\theta_2) \rangle \ , \
 \langle J_- (0) | P_1 (\theta_1) P_2 (\theta_2)
P_3 (\theta_3) P_4 (\theta_4)
\rangle \ , \ \cdots\\
&&\langle U (0) | P_1 (\theta_1)\rangle \ , \
 \langle U(0) | P_1 (\theta_1) P_2 (\theta_2)
P_3 (\theta_3)\rangle \ , \ \cdots
\end{eqnarray}
Symmetry arguments imply that odd (even)
state form factors vanish in the series of eqn.~(4)((5)),
the matrix elements between incoming and
outgoing states are related to these by crossing symmetry and
do not require additional calculations, and relativistic
invariance lets us trivially relate these to $x^-\neq 0$
form factors.  Form factors of series (4) are required
for $LT$ interactions of eqn.~(2). We must insert a complete
set of states between $J(x)$ and $J(y)$.   Form factors
of series (5) are required for $TT$ interactions in eqn.~(3).
For the $SU(3)_L \otimes SU(3)_R$ NLSM, they
have not been explicitly calculated to my knowledge, so the
rest of the lecture will discuss some of the theoretical tools
that help calculate form factors . To simplify the presentation,
I focus on the $SU(2)$ case.

\vskip.1in
{\bf 2. The Integrable $SU(2)$ NLSM}
\vskip.1in
The quantum principle chiral $SU(2)\otimes SU(2)$ NLSM is a completely
integrable system.  This means that it has an infinite number of conserved
currents -- for two dimensional
systems with pure elastic scattering, not
only is momentum conserved, but the entire momentum distribution
is conserved.\footnote{This can be demonstrated on a pool table by lining
up a straight column of touching balls and shooting the cue ball
straight onto the leading ball.  Only one ball comes out the other end.
This is in contrast to the $2+1$ dimensional case which occurs
when one begins the pool game with a `break' of a two dimensional
array of balls.}
This implies that full $S$ matrix factorizes into a product of
two particle S matrices [5].  Symmetry arguments -- unitarity,
crossing symmetry, and the Yang-Baxter equation (the time
ordering of the scattering process is irrelevant to the final S matrix), are
almost sufficient to
determine unique two particle S-matrices.  This is elegantly discussed
in the `Zamolochikov$^2$' paper[6].  The remaining ambiguities are the
initial group multiplets of the particles, and a minimality
assumption on the number bound states.  For the $SU(2)$ NLSM, we assume
that particles are in the fundamental representations
of the left and right global $SU(2)$ symmetries of the action,
and (self consistently) that there are
no additional states.
Each particle is labeled by four states of the $2_L \otimes 2_R$
representation,
which I will parameterize in a matrix notation,
$P  =  P^0 + i\sigma^a P^a$\, ,
$P^\dagger  = P^0 - i\sigma^a P^a$\, ,
where $a = 1,2,3$, and $ \sigma^a$ are the Pauli sigma matrices.
The $P^A$ states, where $A = 0,1,2,3$, transform under global left
symmetry as $[Q_L , P] = i \sigma^a \epsilon^a P$, and
$[Q_L , P^\dagger ] = -i  P^\dagger \sigma^a \epsilon^a$,
or in terms of components,
$\delta_L^a P^0 = - P^a $,
$\delta_L^a P^b = (\delta^{ab} P^0 + \epsilon^{abc} P^c )$.
The transformation
rules under the $SU(2)_R$ group are obtained by exchanging $P$ and
$P^\dagger$.
The two particle S matrix for the $SU(2)$ NLSM was given by
Wiegmann[7], based on the work of Berg, Karowski, Weisz, and Kurak[8].
In the basis of states given above, the S matrix
${\cal S} = S \Sigma$ is
\begin{equation}
P^A (\theta_1) P^B (\theta_2)  = S(\theta )\Sigma^{ABCD}(\theta)\,
P^C (\theta_2) P^D (\theta_1) \ , \ \theta \equiv \theta_1 - \theta_2 \ ,
\label{zf1}
\end{equation}
$S(\theta) =
[\Gamma(- \theta/2\pi i)
\Gamma(1/2 + \theta/2\pi i)]^2/
[\Gamma(\theta/2\pi i)
\Gamma(1/2 - \theta/2\pi i)
(\theta - i\pi)]^2$\, ,
and the factor
$\Sigma^{ABCD}(\theta)  =  ( (i\pi)^2 - i\pi\theta )
\delta^{AC} \delta^{BD}
+ ( \theta^2 - i\pi\theta ) \delta^{AD} \delta^{BC}
+ (i\pi\theta) \delta^{AB} \delta^{CD}$.
As $\theta \rightarrow \infty$ states commute.

The Zamolochikov -- Faddeev (zf) algebra is the full realization of
the S matrix and normalization in terms of creation and
annihilation operators acting on a vacuum state $\invac$.
Asymptotic `in' states are a representation of the S matrix, and
their algebra is given by eqn.~(\ref{zf1}).
Normalization for two particle `in' states is
$\outbrace P^D (\theta_2^\prime ) P^C (\theta_1^\prime ) |
P^A (\theta_1 ) P^B (\theta_2 )\inbrace$  $ = $
$\delta^{AC} \delta^{BD}
\delta (\theta_1 - \theta_1^\prime )
\delta (\theta_2 - \theta_2^\prime )$,
for $\theta_1 > \theta_2$ and $\theta_1^\prime >
\theta_2^\prime$.
We define
$\Pbar^A$ to be annihilation operators satisfying
$\Pbar^A \invac = 0$.
I find the additional self-consistant algebraic relations
\begin{eqnarray}
\Pbar^A (\theta_1) \Pbar^B (\theta_2)  & = & S(\theta )
\Sigma^{ABCD}(\theta) \,
\Pbar^C (\theta_2) \Pbar^D (\theta_1) \ , \label{zf2} \\
\Pbar^A (\theta_1 ) P^B (\theta_2 ) & = &
S( -\theta )\Sigma^{ACBD} (-\theta ) \,
P^C (\theta_2 ) \Pbar^D (\theta_1) + \delta^{AB}\delta(\theta_1
- \theta_2) \ .
\label{zf3}
\end{eqnarray}
Eqns.~(\ref{zf1}, \ref{zf2}, \ref{zf3}) denote the full zf algebra
for the case at hand.  The Fock space of quantum states is
given by an ordered (with respect to rapidity, lesser rapidity
particles to the left is the convention) set of creation
operators acting on the vacuum.
The operator $\rho (\theta) = P^A (\theta ) \Pbar^A (\theta )$ has
simple commutation relations with the particles,
\begin{equation}
[ \rho (\theta_1 ) ,\,  P^A (\theta_2 ) ] =
P^A (\theta_1) \delta (\theta_1 - \theta_2 ) \ , \
[ \rho (\theta_1 ) ,\,  \Pbar^A (\theta_2 ) ] =
-\Pbar^A (\theta_1) \delta (\theta_1 - \theta_2 ) \ .
\end{equation}
With these commutators, it is trivial to construct the light-front
Hamiltonian and momentum:
$P^{\pm} = m \int d\theta  e^{\pm \theta} \rho (\theta ) $\, ,
where $m$ is the mass of the particles.  Clearly, all operators
of the form $\int d\theta e^{n\theta} \rho (\theta)$
are diagonal in this basis of states.  These operators
for integer $n$ correspond to
moments of the light-front momentum of a multi-particle system,
which we expected to exist by assumption.  It is remarkable that they
have such a simple form in this basis of states!

\vskip.1in
{\bf 3. Form Factors}
\vskip.1in

The relationship between the asymptotic particles
and the operators $J$ (and possibly even $U$) which occur
in the interacting transverse lattice Hamiltonian is very complicated.
In the
remaining part of this lecture, I will discuss the first
form factor of eqn.~(5).  Unfortunately, we do not know the exact relation
between the asymptotic states and the current
$J$\footnote{This problem
has an analogy in the Ising model about the critical point,
which has a Lagrangian of free massive fermions.  The
order and disorder operators in the Ising model
are very complicated functions
of the fermion, and their form factors are very non-trivial,
even though the S matrix elements of the fermions are just $\pm 1$.}.
However we are optimistic that these form factors
can be evaluated, because there exist similar result
for the $SU(2)$ Thirring model[9].
I will now construct the exact two point $SU(2)\otimes SU(2)$ form factor
below, using  rules which were applied in the Thirring model case.
The two point form factor for the left current
is defined as
\begin{equation}
e^{+ i x^- [ p^+ (\theta_1) + p^+ (\theta_2)]} \,
\langle  J^a_L (x^-) P^B (\theta_1 ) P^C (\theta_2 )  \rangle =
\langle  J^a_L (0) P^B (\theta_1 ) P^C (\theta_2 )  \rangle  =
\FF (\theta_1, \theta_2)
\label{formfactor}
\end{equation}
where the amplitude is taken to be between asymptotic `in' and `out' vacua.
Crossing symmetry implies that if $\theta_1 \rightarrow \theta_1 + i\pi$
in the amplitude, the $P^B$ `in' field becomes an `out' field:
$\FF (\theta_1 + i\pi, \theta_2 ) =  \langle \Pbar^B (\theta_1)  J^a_L (0)
P^C (\theta_2 )  \rangle$\, .
The current is assumed to be conserved, with normalization
\begin{equation}
\int dx^- \langle \Pbar^B (\theta_1 ) J^a_L (x^- ) P^C(\theta_2 )
\rangle = \langle \Pbar^B(\theta_1) Q^a_L P^C(\theta_2 ) \rangle =
{\rm tr} [ \sigma^a T^{BC}_L ] \delta(\theta_1 - \theta_2 ) \ .
\label{normal}
\end{equation}
By comparing to the left transformations on the particles, we find
the $T$ matrix to be antisymmetric with matrix elements
$T^{00}_L = 0$\, ,
$T^{0b}_L =\sigma^b / 2$\, ,
$T^{bc}_L = - \epsilon^{bcd}\sigma^d / 2$\, .
Note that ${\rm tr} [\sigma^a T^{BC}_L]$ is simply the Clebsh-Gordon
coefficient that corresponds to the left triplet channel in
the product of two left-handed doublets.

The form factor must satisfy an additional
set of reasonable constraints, the {\bf Smirnov
Axioms}[10].  In the context of the two point form factor, they can be
stated as:
\underline{Axiom 1:}  The function $\FF (\theta_1, \theta_2)$ is
analytic in $\theta$ in the strip
$0\le \Im \theta \le 2\pi$, except for single poles.
It becomes the physical form factor for
real $\theta_i$ and $\theta_1 < \theta_2$.
The poles are on the imaginary axis and of of two types,
either the fusion of physical states onto intermediate physical
bound states, or  particle - anti-particle fusion onto the vacuum.
\underline{Axiom 2:}
Relativistic invariance requires that the form factor satisfies
$\FF (\theta_1 +\beta, \theta_2 + \beta)$ $=$
$\exp{[s(J)\beta]}\FF (\theta_1,\theta_2)$,
where $s(J) = -1$ is the spin of $J_- $ in our case.
\underline{Axiom 3:}
The form factor should satisfy the symmetry property (Watson's Theorem)
$\FF (\theta_1, \theta_2) = \Sigma^{BCDE} (\theta_1 - \theta_2)
S(\theta_1 - \theta_2)
F^{DE}_{J^a_L} (\theta_2, \theta_1)$.
This seen by applying eqn.(\ref{zf1}) to the
form factor, or equivalently by inserting a complete set of states
between $P^B$ and $P^C$.
\underline{Axiom 4:}
The form factor satisfies
$ \FF (\theta_1, \theta_2 + 2\pi i) = e^{2\pi i w(J,U)}F^{CB}_{J^a_-}
(\theta_2, \theta_1)$,
where $w(J,U)$ is the relative monodromy index
between the current and the fundamental field corresponding to $U$.
This corresponds to a Euclidean continuation of one operator about the
other, and $w = 0$ in our bosonic case.
The ansatz for the form of $\FF$ which satisfies these
axioms is
\begin{equation}
\FF (\theta_1, \theta_2) = - \frac{1}{4\pi}[p^- (\theta_1 ) +
p^- (\theta_2 )] \, {\rm Tr}[\sigma^a T^{BC}_L] \tanh (\theta/2)
 F(\theta) \ .
\label{ansatz}
\end{equation}
The pole at $\theta = i\pi$ in the $\tanh$ function satisfies
axiom 1 for the scalar part of the form factor
and corresponds to the fusion of the two doublets
onto the vacuum state.  For the full vector form factor,
this pole is cancelled by the zero in $p^- (\theta_1 ) + p^- (\theta_2 )$,
giving a finite result at the normalization point $\theta = i\pi$,
and the normalization condition eqn.~(\ref{normal}) is satisfied
when $F(i\pi ) = 1$.
The zero in the $\tanh$ function
at $\theta = 0$ exists by axiom 3 since the S-matrix in this channel is
$-1$ at $\theta = 0$.  Eqn.~(\ref{ansatz}) satisfies relativistic
invariance (axiom 2) as long as $F = F(\theta_1 - \theta_2)$.
$F$ has no poles or zeros in the region
$0\le \Im \theta \le 2\pi$, and it satisfies
\begin{equation}
8\pi i \frac{d}{d\theta} F(\theta ) =
\int_C \frac{dz}{\sinh^2 {\frac{1}{2} (z - \theta)}} \ln F(z)
=
\int^\infty_\infty \frac{dz}{\sinh^2 {\frac{1}{2} (z - \theta)}}
\ln {\frac{F(z)}{F(z+2\pi i)}} \ .
\end{equation}
where the contour $C$ goes from $[-\infty, \infty ]$
up to the $2\pi i $ line,
from
$[\infty +2\pi i, -\infty + 2\pi i ]$,
and is closed by going back to  the real line.  Assuming that
the form factor falls off faster than $\ln F \rightarrow e^{z}$
(which we will verify below), we
can ignore the end caps to get the second equality above.
{}From axiom 3, $F(\theta) = \bar{S} (\theta) F (-\theta )$
and from axiom 4, $F(\theta + 2\pi i ) = F(-\theta )$, so
\begin{equation}
\frac{F(\theta)}{F(\theta + 2\pi i)} =
S(\theta)(\theta - i\pi )(\theta+i\pi )
\equiv \bar{S}(\theta) =
\frac{\Gamma^2 (1 -\topi )\Gamma (\frac{1}{2} + \topi )\
\Gamma (\frac{3}{2} + \topi)}
{\Gamma^2 (1 + \topi )\Gamma (\frac{1}{2} - \topi )\
\Gamma (\frac{3}{2} - \topi)} \ .
\end{equation}
If $\bar{S}(\theta)  = \exp {\int^\infty_0 \frac{dx}{x} f(x)
\sinh{\frac{x\theta}{i\pi}} }$\ ,
then
$F(\theta) = \exp \int^\infty_0 \frac{dx}{x} f(x)
\sin^2 [x(i\pi - \theta )/2\pi]/\sinh x $~[11].
For the case at hand, we find
$f(x) = - (1 - e^{-x})^2 / \sinh x$.
For large $Q^2 = m e^\theta$, the result simplifies and one finds
$F(\theta ) = [\ln \frac{Q^2}{m^2}]^{-1/2}$\, .
This is the expected asymptotic freedom of the $SU(2)$ NLSM\footnote{
In the transverse lattice picture, this form factor for large
$Q^2$ is a dressed three point transverse gluon vertex.
The above result means that the
sum over a certain class of transverse gluons renormalizes the
charge such that $\alpha (Q^2 ) \sim 1/[\ln \frac{Q^2}{m^2}]$.
}.
Higher point form factors are more complicated because of the number
of possible group theoretic ways the states can factor onto the
current.  Recently, a bosonization prescription has been given for
the construction of higher point form factors in the $SU(2)$ Thirring
model case[12].

We clearly need to compute
more NLSM form factors before full computer simulations can begin.
With the connection between the $SU(3)\otimes SU(3)$ NLSM and
QCD made explicit by the transverse lattice, these
calculations take on new phenomenological relevance.
This work was partly supported by Komitet Bada{\'n} Naukowych
Grant KBN-2P302-031-05, SSC fellowship FCFY9318,
and DOE grant DE-FG05-86ER-40272.

\vskip.1in
{\bf References}
\vskip.1in

1.~A.~Casher, Phys.~Rev.~D14 (1975) 452.  C.~Thorn, Phys.~Rev.~D19 (1976) 94.

\ \ \ H.C.~Pauli and S.J.~Brodsky, Phys.~Rev.~D32 (1987) 1993.

2.~M.~Burkardt, Nucl.~Phys.~A504 (1989) 762.
K.~Hornbostel, S.~J.~Brodsky (SLAC),

\ \ \ H.~C.~Pauli, Phys.~Rev.~D41 (1990) 3814.

3.~M.~Burkardt, Phys.~Rev.~D49 (1994) 5446, hep-th/9312006.

4.~W.~Bardeen, R.~Pearson, Phys.~Rev.~D14 (1976) 547.
W.~Bardeen, R.~Pearson,

\ \ \ E.~Rabinovici, Phys.~Rev.~D21 (1980) 1037.

5.~K.~Pohlmeyer, Comm.~Math.~Phys.~46 (1976) 207.
A.~Polyakov, Phys.~Lett.~72b

\ \ \ (1977) 224.
Y.Y.~GoldSchmidt and E.~Witten, Phys.~Lett.~91b (1980) 392.

6.~Al.~B.~Zamolodchikov and Ale.~B.~Zamolodchikov, Ann.~Phys.~120
(1979) 253.

7.~P.~Wiegmann, Phys.~Lett.~142b (1984) 173.

8.~B.~Berg, et.~al.~Nucl.~Phys.~B134 (1978) 125. (See table 1, Case II.)

9.~A.N.~Kirillov and F.A.~Smirnov, Phys.~Lett.~198b (1987) 506.

10.~F.A.~Smirnov, J.~Phys.~A 19 (1986) L575.

11.~M.~Karowski and P. Weisz, Nucl.~Phys.~B139 (1978) 455.

12.~S.~Lukyankov, Rutgers Preprint RU-93-30 (1993), hep-th/9307196.
\end{document}